\documentclass[a4paper,11pt]{article}
\pdfoutput=1 

\usepackage{jcappub} 

\usepackage[T1]{fontenc} 

\newcommand{\edit}[1]{\textcolor{black}{{} #1}}

\begin{document}

\title{Constraints on the dark matter-baryon interaction cross section from galaxy cluster thermodynamics}


\author[1]{E. Stuart,\note{Corresponding author.}}
\author{K. Pardo,}


\affiliation{%
 University of Southern California \\
825 Bloom Walk\\
Los Angeles, CA 90089-1483
}%

\emailAdd{stuarte@usc.edu}

\abstract{Dark matter (DM) models with a non-zero DM-baryon interaction cross section imply energy transfer between DM and baryons. We present a new method of constraining the DM-baryon interaction cross section and DM particle mass for velocity-independent interactions using the thermodynamics of galaxy clusters. If the baryonic gas in these clusters is in thermodynamic equilibrium and DM cools baryons, this cooling rate is limited by the net heating rate of other mechanisms in the cluster. We use the REFLEX clusters from the Meta-Catalogue of X-ray detected Clusters of Galaxies (MCXC) with mass estimates from the Atacama Cosmology Telescope (ACT) catalog of Sunyaev-Zel'dovich (SZ) selected galaxy clusters. This yields 95\% upper bounds on the DM-proton interaction cross section for velocity-independent interactions of $\sigma_0\leq9.3\times10^{-28} \mathrm{~cm^2}$ for DM masses, $m_\chi = 10^{-4} - 10^{-1}$ GeV. These constraints are within an order of magnitude of the best constraints derived in this mass range, and serve as a complementary, independent constraint. We also apply this model to the fractional interacting DM scenario, where only 10\% and 1\% of the DM is interacting. Unlike other methods, this constraint scales \textit{linearly} with this fraction. This yields 95\% upper bounds of $\sigma_0\leq1.1\times10^{-26} \mathrm{~cm^2}$ and $\sigma_0\leq8.2\times10^{-26} \mathrm{~cm^2}$, which are the strongest existing constraints for this scenario. This paper serves as a proof of concept. Upcoming SZ measurements will provide temperature profiles for galaxy clusters. Combining these measurements with more complex thermodynamic models could lead to more robust constraints. 
}

\arxivnumber{2411.18706}

\maketitle
\flushbottom

\section{Introduction} \label{sec:intro}

Many astrophysical and cosmological observations indicate that dark matter (DM) makes up the majority of matter in the Universe, but there are still many open questions about its properties. The fiducial model of DM, Cold Dark Matter (CDM), where DM is non-relativistic and non-interacting, satisfies most observations but leaves many questions to be answered. For example, the mass of the DM particle and the value of any interaction cross sections remain unknown. Some models of interest include DM-baryon interactions.

There are several existing constraints on the DM-baryon cross section for velocity independent interactions. For example, Refs.~\cite{Gluscevic2018, Ali-Haimoud2015, Xu2018} examine how these interactions would be expected to alter our observations of the cosmic microwave background (CMB), excluding significant parameter space for DM-proton scattering. Ref.~\cite{Maamari2021} uses measurements of the MW satellite population from the Dark Energy Survey and Pan-STARRS1 to improve on these constraints. Ref.~\cite{Wadekar2021} uses the gas cooling rate in dwarf galaxies, and in particular the Leo T dwarf galaxy, to constrain these interactions. However, significant parameter space remains to be explored, particularly for cases where the interacting DM makes up only a fraction of the total DM.

Galaxy clusters, composed of mostly DM and ionized hydrogen gas, are the largest gravitationally bound structures in the universe. These properties make them a unique environment for observing phenomena such as structure formation, galaxy-galaxy interactions, and the relationship between DM and baryon distributions. X-ray observations provide the luminosity of clusters, which can be converted to other properties such as mass and temperature through scaling relations \citep{Gaspari2019}. Additionally, the Sunyaev-Zel’dovich Effect (SZE) \citep{Sunyaev1972}, a distortion of the CMB caused by photons scattering off of the gas in a cluster, gives insight into cluster thermodynamics. The CMB is precisely measured, so these distortions can be used to calculate the temperature and density profiles of clusters \citep{Amodeo2021, Schaan2021}, as well as estimate mass through a scaling relation \citep{Hilton2021}. These measurements are new, and have only been detected by stacking analyses, but will continue to improve through surveys from the upcoming Simons Observatory \citep{Ade2019}. By combining multiple cluster measurements, we can learn details about galaxy clusters and their DM.

Previous work has investigated the potential for DM and baryons to exchange energy through different types of interactions \citep{Munoz2015, Shoji2024, Qin2001, Wadekar2021}. For example, ref.~\cite{Shoji2024} investigated the possibility for velocity independent DM-baryon interactions to explain the cluster cooling flow problem \citep{Cavagnolo2009, Peterson2003}, assuming heating due to DM-baryon interactions. We consider the alternative, that interactions with DM particles cool the baryons. We consider galaxy clusters assumed to be in thermal equilibrium. The baryons are heated via active galactic nuclei (AGN) feedback and cooled via radiative cooling (bremsstrahlung). The effects of cooling due to DM-baryon interactions are then constrained by the difference in magnitude of these two mechanisms.  
 
This paper aims to further constrain the DM-baryon interaction cross section and DM particle mass by exploring the gas thermodynamics of galaxy clusters. It is organized as follows: in section~\ref{sec:modeling} we explain the thermodynamic model of galaxy clusters and DM-baryon interactions used. In section~\ref{sec:data} we discuss the dataset we use to test the model. In section~\ref{sec:results} we discuss the constraints obtained by applying this model to the dataset, and in section~\ref{sec: discussion} we discuss the implications of these results and how upcoming measurements will improve our model. 

\section{Modeling} \label{sec:modeling}

This section details the models of heating and cooling within galaxy clusters and how these models can be combined to produce a constraint on the DM-baryon interaction cross section. We present the models for radiative cooling and heating due to AGN feedback in a galaxy cluster, define the model for heat transfer through baryon-DM interactions, and show how these interactions are constrained by an assumption of thermal equilibrium.


Ref.~\cite{Sereno2021} found that galaxy clusters are approximately virialized by $z\sim0.2$. The thermal kinetic energy and gravitational potential energy in a cluster with an ideal hydrogen gas are \citep{Sereno2021}
$E_{th} = \frac{3}{2}k_B T_b$, and $E_{p} = -\frac{3}{2} \frac{G \mu m_p M_{500}}{R_{500}}$, where $R_{500}$ is the radius within which the average density of the cluster is 500 times the critical density, $M_{500}$ is the mass within $R_{500}$, mean molecular weight $\mu=0.59$, $G$ is the gravitational constant, $k_B$ is the Boltzmann constant, and $m_p$ is the mass of a proton.\footnote{\edit{We consider the mass within $R_{500}$ because beyond this radius accretion rates play a larger role and it is not safe to assume the gas is mostly virialized \cite{Sereno2021}.}} In general, a cluster's mass is either nearly constant or increases as it accretes material from its surroundings \citep{Sereno2021}. Therefore, the magnitude of its potential energy also remains constant or increases. For clusters to remain virialized, their kinetic energy and therefore their temperatures must also be constant or increasing. To maintain virial equilibrium, the combined effect of heating and cooling processes must therefore result in a net heating rate $\geq 0$. Thus, an upper limit on a cluster's cooling rate can be determined by assuming thermal equilibrium.

Ref.~\cite{Shaw2006} point out that because galaxy clusters are not isolated systems, the virial theorem should be corrected with an additional term that represents surface pressure, so $2E_{th} + E_{p} - E_{sp} = 0$, where $E_{sp}$ is the average effect of surface pressure on a particle, $E_{sp}=\frac{\mu m_p}{M_{500}}E_s$ with \edit{$E_s \approx 4\pi R^3_{0.9}P_s$, where $R_{0.9}$ is the median radius of the outermost 20\% of the particles in the halo, and $P_s$ is the pressure at the boundary of the halo} as in ref.~\cite{Shaw2006}. Including this additional surface pressure term has the effect of changing the exact subset of clusters that are virialized. However, as we briefly discuss in section~\ref{sec:results}, this does not substantially change the results presented in this paper.

To ensure cooling due to DM-baryon interactions does not disrupt the thermal equilibrium of baryons in the cluster, the combined effects of the heating rate in a cluster due to AGN feedback, the cooling rate due to bremsstrahlung emission and the rate baryons lose energy to DM should be approximately equal to zero net heating:
\begin{equation}
\label{equilibrium condition}
    \frac{dQ_{\mathrm{AGN}}}{dt} + \frac{dQ_{\mathrm{rad}}}{dt} + \frac{dQ_{\mathrm{b\rightarrow\chi}}}{dt}\approx 0,
\end{equation}
where $\frac{dQ_{\mathrm{AGN}}}{dt} $ is the heating rate due to AGN feedback, $\frac{dQ_{\mathrm{rad}}}{dt}$ is the heat transfer rate due to bremsstrahlung, and $\frac{dQ_{\mathrm{b\rightarrow\chi}}}{dt}\approx 0$ is the heat transfer rate due to baryons losing energy to DM particles. Since  $\frac{dQ_{\mathrm{AGN}}}{dt} $ and $\frac{dQ_{\mathrm{rad}}}{dt}$ are similar in magnitude, this places an upper bound on the magnitude of the cooling rate due to DM-baryon interactions, as those interactions cannot cool the cluster more than is allowed by this equilibrium statement.

Calculating heating and cooling rates for galaxy clusters requires information about their gas densities and temperatures, \edit{as our only observables are $M_{500}$ and $L_{500}$ (section  \ref{sec:data} for more details)}. The density and temperature profiles of baryons in the intracluster medium (ICM) is determined using the hydrostatic equilibrium equation assuming spherical symmetry. 
\begin{equation}
\label{eq: hydrostatic}
    \frac{dP_b}{dr} = -\rho_b \frac{G M_{\mathrm{enc}}(r)}{r^2}
\end{equation}
where temperature is related to density $\rho_b$ through the ideal gas law $T_b = \mu m_p P_b/\rho_b$. \edit{ $M_{\mathrm{enc}}(r)$ is the total mass enclosed within a radius $r$, assuming the Navarro-Frenk-White (NFW) density profile \citep{Navarro1996}, and can be expressed as
\begin{equation}
\label{eqn:menc}
    M_{\mathrm{enc}}(r) = 4\pi r_s^3 \rho_s \left(\mathrm{ln}(1+r/r_s) - \frac{r/r_s}{1+r/r_s}\right),
\end{equation}
where $r_s$ is the scale radius and $\rho_s$ is the normalization of the density profile.} \edit{$P_b$ is the ICM pressure profile defined  as
\begin{equation} \label{eqn Pb}
    P_b(x) = \frac{P_0 P_{500}}{(c_{500}x)^{\gamma}[1+(c_{500}x)^{\alpha}]^{(\beta-\gamma)/\alpha}},
\end{equation} 
where $x=r/R_{500}$, $P_{500}$ is a function of mass and redshift, and $P_0$, $c_{500}$, $\gamma$, $\alpha$, and $\beta$ are model parameters listed in ref.~\cite{Iqbal2023}. }

Ref.~\cite{Gaspari2019} found the relationship between temperature and luminosity 
as $\mathrm{log}\left(\frac{L_{500}}{10^{44}~\mathrm{erg/s }}\right) = (-2.34 \pm 0.09) + (4.71 \pm 0.26)\mathrm{log}\left(\frac{T_c}{\mathrm{keV}}\right)$, where $T_c$, the core temperature, is stated to be comparable to $T_{500}$, the temperature at $R_{500}$, and $L_{500}$ is the X-ray luminosity within $R_{500}$. 
We normalize the temperature profile obtained through the hydrostatic equilibrium equation (eq.~\ref{eq: hydrostatic}) so $T_{500}$ is consistent with the ref.~\cite{Gaspari2019} relationship, \edit{in order to get a radially dependent temperature profile where the core temperature is hotter than the virial temperature}. 

Baryons in the ICM of galaxy clusters have temperatures of about $10^6$--$10^8$ K and lose energy through bremsstrahlung emission \citep{Rybicki1979}. We take the Gaunt factor $g(\nu,T)=1$, valid for the temperature of galaxy clusters \citep{Mo2010}, and assume the same number density of electrons and protons: $n_e=n_i=n_b$, for an ICM made of mostly ionized hydrogen. We also take charge number $Z^2=1$ for ionized hydrogen, which provides an underestimate of bremsstrahlung emission, leading to more conservative constraints. Integrating the bremsstrahlung emissivity over all frequencies and the volume of the cluster gives the total radiative cooling rate due to bremsstrahlung as 
\begin{equation}
\label{eqn cooling rate}
    \frac{dQ_{\mathrm{rad}}}{dt} = - \int_{R_{\mathrm{min}}}^{R_{500}} 6.8 \times 10^{-42}\frac{n_b(r)^2}{T_8(r)^{1/2}} \frac{k_BT_b(r)}{h} 4\pi r^2dr,
\end{equation}
where $T_8(r)=T_b(r)/10^8$ K. For the results presented in this paper, we use $R_{\mathrm{min}} = 0.001R_{500}$.

AGN feedback is an active area of research, with many different heating mechanisms proposed, such as heating proportional to chaotic cold accretion onto supermassive black holes \citep{Gaspari2019}, and radio jets \citep{Binney1995}. We use the effervescent heating model from ref.~\cite{Iqbal2023} to estimate the heating rate due to AGN feedback in a cluster. This model describes bubbles that are created by AGN and get carried outwards in the cluster due to a pressure gradient in the ICM. The heating rate for this model is
\begin{equation}
\label{eqn heating rate}
    \frac{dQ_{\mathrm{AGN}}}{dt} \approx \int_{R_{\mathrm{min}}}^{R_{500}} h(r)P_b(r)^{(\gamma_b-1)/\gamma_b}\frac{1}{r}\frac{d\mathrm{ln}P_b(r)}{d\mathrm{ln}r} 4\pi r^2dr \; ,
\end{equation}
where $h(r)$ is a term accounting for the magnitude and radial dependence of the heating rate defined in appendix \ref{app:agn model}. This AGN heating model was selected because, unlike some other numerical models for AGN (such as the chaotic cold accretion in ref.~\citealp{Gaspari2019}), it does not directly depend on the cooling rate/cooling time of the cluster. Instead, the magnitude of the AGN heating rate is a function of the central black hole mass (see appendix \ref{app:agn model}). Therefore, the heating rate can be considered independent of the cooling rate used in this work. The predictions made by this model are consistent with existing scaling relations \citep{Iqbal2023}.

For interactions between baryons and DM with velocity-dependent interaction cross sections like $\bar{\sigma} = \sigma_0v^n$, the expression for the cooling rate due to baryons losing energy to DM particles is as follows \citep{Munoz2015, Shoji2024}: 
\begin{equation}
\label{eqn DM cooling}
        \frac{dQ_{b\rightarrow\chi}}{dt} = \int_{R_{\mathrm{min}}}^{R_{500}} \frac{3(T_{\chi}(r)-T_b(r))\rho_{\chi}(r)\rho_b(r)\sigma_{0}c_n u_{th}(r)^{n+1}}{(m_{\chi} + m_{b})^{2}} 4\pi r^2dr\; ,
\end{equation}
where $u_{th}(r) = \left(\frac{T_{\chi}(r)}{m_{\chi}}+\frac{T_b(r)}{m_b}\right)^{1/2}$, $n$ represents different types of velocity-dependent interactions (for example, $n=0$ for collisions, $n=-2$ for dipole interactions, and $n=-4$ for Coulomb-like interactions), and $c_n=\frac{2^{(5+n)/2}}{3\sqrt{\pi}}\Gamma(3+n/2)$. We include a brief summary of the derivation for this equation in appendix \ref{app:dm-baryon}. In this work, we focus on $n=0$ (velocity independent) interactions. $T_b>T_\chi$ is required for baryons to lose energy through these interactions. \edit{The expected dependence of $\sigma_0$ on $m_\chi$ can be seen by solving eq.~\ref{equilibrium condition} for $\sigma_0$.
\begin{equation}
\label{sigma_0 expression1}
        \sigma_0 = \left(\frac{dQ_{AGN}}{dt} + \frac{dQ_{rad}}{dt} \right)\frac{(m_{\chi} + m_{b})^{2}}{3(T_{b}-T_\chi)\rho_{\chi}\rho_b u_{th}Vc_0} \; .
\end{equation}}

The temperature of DM follows from the virial theorem: 
$T_\chi(r) \approx 0.3 \frac{GM_{\mathrm{enc}}(r)}{r}m_\chi$,
where $m_\chi$ is the mass of the DM particle. This gives a radially dependent temperature profile which is equal to the virial temperature at the virial radius, and higher at lower radii. Some of the kinetic energy of the clusters is in baryons, therefore this yields an overestimate of $T_\chi$, producing more conservative constraints. We also ignore surface pressure in this calculation, which again leads to an overestimate of $T_\chi$. Our model requires $T_\chi<T_b$ for cooling due to DM-baryon interactions. The equation satisfies this requirement as long as $m_\chi < m_p$ in a model where 100\% of the DM has interactions with baryons. 


If the interaction rate between baryons and DM is too high, the interactions will thermalize the DM and baryons at an earlier time, so no further cooling would occur today. To ensure DM cooling will not thermalize the cluster over its lifetime ($\sim$10 Gyr), we verify that:
\begin{equation}
\label{eqn thermalization}
    \frac{1}{\Gamma} = \frac{1}{n_\chi \sigma_0 v} > 10 \ \mathrm{Gyr}
\end{equation}
where $v=(GM_{500}/R_{500})^{1/2}$ is the velocity of a particle. This inequality defines a region within which the method presented in this paper cannot be applied. Within that region, DM-baryon interactions are too frequent to approximate that $T_b$ and $T_\chi$ are constant with $T_b>T_\chi$. This condition is more limiting for smaller $m_\chi$ because of the dependence on number density, $n_\chi$. In our analysis, we take the average of the line generated by this inequality for all clusters and use that to define the prior. We also only perform this analysis for $m_\chi<0.1 \mathrm{~GeV}$ because above this mass, $T_\chi$ starts approaching $T_b$ at some radii of the cluster. $T_\chi > T_b$ would indicate heating due to DM-baryon interactions, as presented in ref.~\cite{Shoji2024}.

\section{Data \& Methods} \label{sec:data}

Estimating the cooling rate due to DM-baryon interactions requires knowledge of the temperatures and densities of DM and baryons. The same information is needed for the cooling due to bremsstrahlung, and injected luminosity is needed to estimate the heating rate due to AGN feedback. These quantities are derived from measurements of $L_{500}$ and $M_{500}$, the mass and luminosity within $R_{500}$, the radius where the density of the cluster is equal to 500 times the critical density of the Universe. 

The ROSAT-ESO Flux Limited X-ray (REFLEX; \citealp{Bohringer2004}) survey reports the measured X-ray flux $F_x$ and percent error, the observed X-ray luminosity  $L_{obs}$, and the luminosity corrected for flux missed outside the detection aperture (7 arcmin or half the virial radius if larger than 7 arcmin) $L_{cor}$ for 447 clusters. The Meta-Catalogue of X-ray detected Clusters of galaxies (MCXC, \citealp{Piffaretti2011}) calculates $L_{500}$ based on $L_{cor}$ for 441 of the clusters in the REFLEX dataset.

Independent mass ($M_{500}$) and luminosity ($L_{500}$) measurements are necessary when determining which clusters are virialized, as the thermal energy is related to luminosity through temperature, and the potential energy is related to mass. To derive our constraints, we need an independent measurement of $M_{500}$ to hold constant. We use the ACT catalog of SZ-selected galaxy clusters \citep{Hilton2021} for its reported $M_{500}$ derived from the SZ signal. This catalog includes $M_{500}$ measurements derived through a $Y_{500}-M_{500}$ scaling relation (where $Y_{500}$ is the compton-Y parameter, an SZ observable) for 225 REFLEX clusters which were matched to the REFLEX dataset through their REFLEX IDs. All of these clusters have reported error bars on masses defined by the 68\% confidence interval that are less than 20\% of their reported masses.

\begin{figure*}
    \centering
    \includegraphics[width=0.9\textwidth]{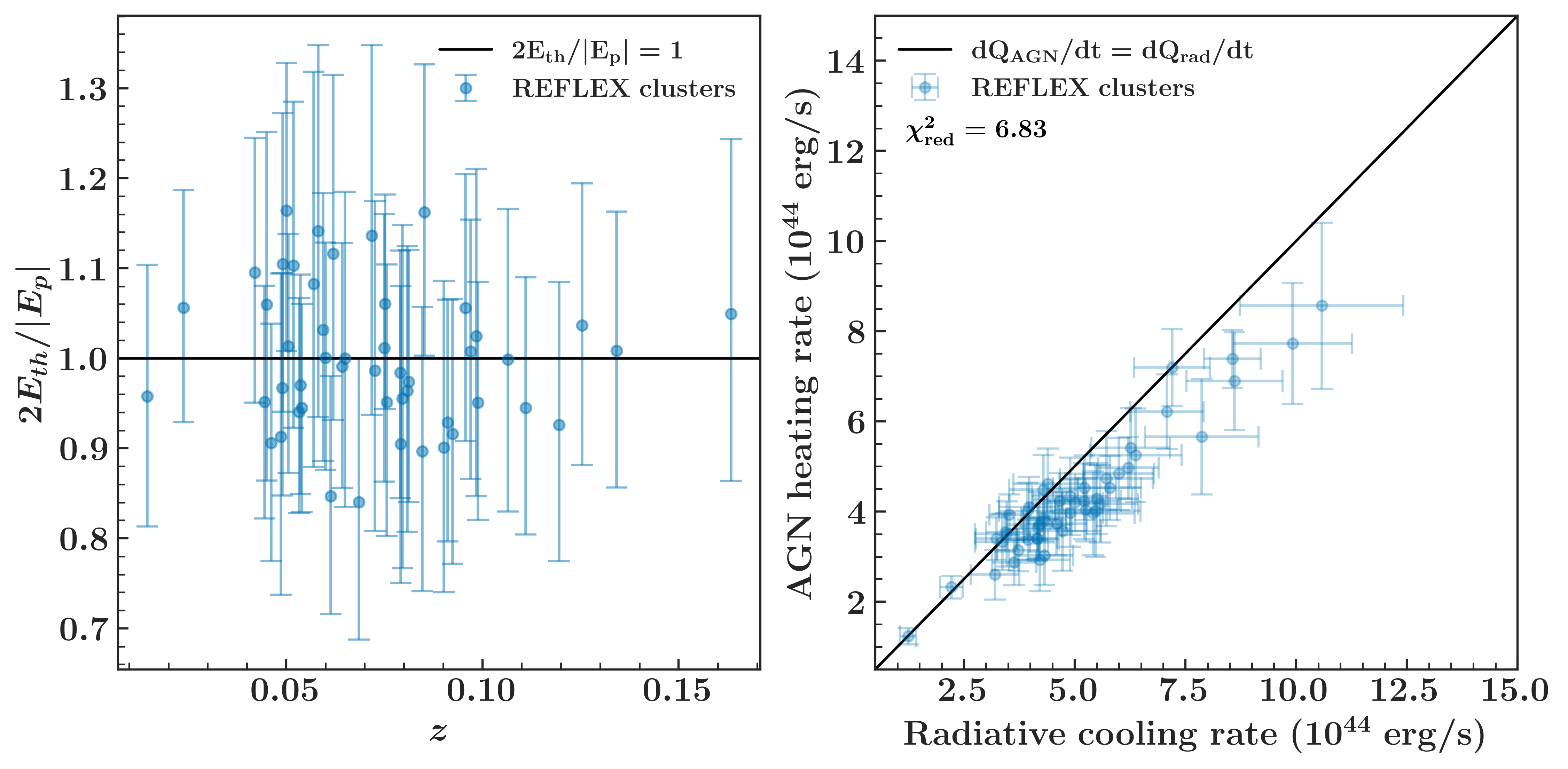}
    \caption{\textbf{Left:} The ratio of 2 times the thermal energy of the ICM to the absolute value of the potential energy for 49 clusters in the REFLEX dataset as a function of redshift. The plotted clusters are all within 2$\sigma$ of the solid black line, which indicates virialization. The reduced chi-squared value $\chi^2_{red}=0.24$ suggests a good fit, indicating the data points align with the virial equilibrium model. \edit{The error bars shown are propagated from the 68\% confidence intervals for $M_{500}$ and $L_{500}$, described in more detail in section \ref{sec:data}.}
    \textbf{Right:} The heating rate due to AGN feedback of the ICM in the same REFLEX clusters based on the ref.~\cite{Iqbal2023} effervescent heating model, vs the cooling rate from bremsstrahlung given in ref.~\cite{Rybicki1979}. The black line represents the line where the AGN heating rate and the radiative cooling rate due to bremsstrahlung are equal in magnitude. The relatively high $\chi^2_{red}=6.83$ and deviation from the black line indicates this model may not fully capture all heating and cooling mechanisms in clusters (see section~\ref{sec: discussion} for more information).}
    \label{fig:equilibrium}
\end{figure*}

We limit our sample to the 173 clusters with $z<0.2$, to be consistent with the redshift by which clusters are considered to be in thermal equilibrium \citep{Sereno2021}. \edit{Poisson 68\% errors are given for $M_{500}$, with a mean positive error of 9.0\%, and mean negative error of 9.4\%. We estimate the $L_{500}$ error using a method described below, which yields an average error of 24\%. From these observables, all other quantities used in calculations are derived. The error bars are propagated through the calculations using the asymmetric\_uncertainty Python package \citep{Gobat2022}.} Then, we reduce the sample to only include clusters within $2\sigma$ of having $2E_{th}/|E_p| = 1$, indicating virialization (see figure \ref{fig:equilibrium}). This leaves a set of 111 clusters, where the removed clusters were usually too cool ($2E_{th}<<|E_p|$). Haloes with early formation times are more likely to be relaxed by late times \citep{Sereno2021}. These virialized clusters are not expected to be undergoing major mergers, but rather evolve through smooth mass accretion. Therefore they should not experience significant shock heating. Then, as our final cut, we remove clusters with AGN heating rates more than $2\sigma$ different from their radiative cooling rates, therefore leaving only 49 clusters that are approximately in thermal equilibrium under this model. \edit{These 49 clusters are shown in figure \ref{fig:equilibrium}.} Performing the analysis described in this paper on the previous set of 111 clusters, or on the set of 173 $z<0.2$ clusters yields results within an order of magnitude of the ones presented in this paper.

Error bars on the observed $L_{500}$ are needed to obtain results using the model proposed in section \ref{sec:modeling}, but are not provided in the MCXC or REFLEX datasets. Given the percent error on the flux measurements in the $z<0.2$ clusters in the REFLEX dataset, we propagate the error to $L_{\mathrm{obs}}$ using the known relationship, $L_{\mathrm{obs}} = 4 \pi d_L^2 F_x$. Then, to estimate the uncertainty of the calculated $L_{500}$ values, we run a Markov Chain Monte Carlo (MCMC) sampler to find best fit parameters $m, b$ such that $L_{500} = mL_{\mathrm{obs}} + b$, and then propagate the variances of those parameters and of $L_{\mathrm{obs}}$ to estimate the uncertainty of $L_{500}$. The MCMC estimates of $m, b$ were $m\approx0.9748\pm4.213 \times 10^{-5}$  and $b\approx 4.704\times10^{37}\pm3.033\times10^{32} \mathrm{~W}$. 

To obtain upper bounds on $\sigma_0$ as a function of $m_\chi$, we plug in the expressions for radiative cooling, AGN heating, and DM cooling (eqs.~\ref{eqn cooling rate}, \ref{eqn heating rate}, \ref{eqn DM cooling}) into the equilibrium condition (eq.~\ref{equilibrium condition}). Then, $T_b$ and $L_{\mathrm{inj}}$ are written in terms of $L_{500}$ through eqs.~\ref{eqn-l500mbh}, and \ref{eqn Linj-Mbh}, and the $L_{500}-T_c$ scaling relation from ref.~\cite{Gaspari2019} to obtain an equation that can be numerically solved for $L_{500}$. We estimate the uncertainty of the theoretical $L_{500}$ to be equal to the magnitude of $L_{500}$. A prior is determined based on the thermalization condition (eq.~\ref{eqn thermalization}) so that we only examine pairs of $\sigma_0$ and $m_\chi$ for which DM cooling is possible. Then we sample the $\sigma_0$, $m_\chi$ using a MCMC sampler, implemented in emcee \citep{emcee} to find the parameters that minimize the $\chi^2$ difference between the observed $L_{500}$ as reported in REFLEX, and the model predicted $L_{500}$ that comes from numerically solving for $L_{500}$ in eq.~\ref{equilibrium condition}. 

\edit{ 
Our MCMC analysis used a uniform prior for $\mathrm{log}(\sigma_0)$ between -40 and -23, and $\mathrm{log}(m_\chi)$ between -4 and -1. These 2 free parameters are independent. The region above the boundary defined by eq. \ref{eqn thermalization} is excluded from this analysis, since above this line interactions are frequent enough that $T_b$ and $T_\chi$ cannot be considered constant and independent. This is represented in figure \ref{fig:constraints comparison}, where the excluded regions have a wedge shape. We initialize 64 chains randomly within the parameter space, allow them to run for a burn-in period of 1000 steps, and then run each of them for 40,000 steps. The log-likelihood function within the prior region is defined as
\begin{equation}
    \mathrm{log}(\mathcal{L}) = -\frac{1}{2}\frac{(L_{\mathrm{500,pred}}-L_{\mathrm{500,obs}})^2}{L_\mathrm{var,pred}^2 +L_\mathrm{var,obs}^2 },
\end{equation}
where $L_{\mathrm{500,pred}}$ is obtained by numerically solving eq. \ref{equilibrium condition} for $L_{500}$, given sampled values for $\sigma_0$ and $m_\chi$, we conservatively estimate $L_\mathrm{var,pred} = L_{\mathrm{500,pred}}$, and $L_{\mathrm{500,obs}}$ and its variance $L_\mathrm{var,obs}$ are the data and variance for $L_{500}$, which we describe earlier in this section. }

\section{Results} \label{sec:results}

\begin{figure*}
    \centering
    \includegraphics[width=0.9\textwidth]{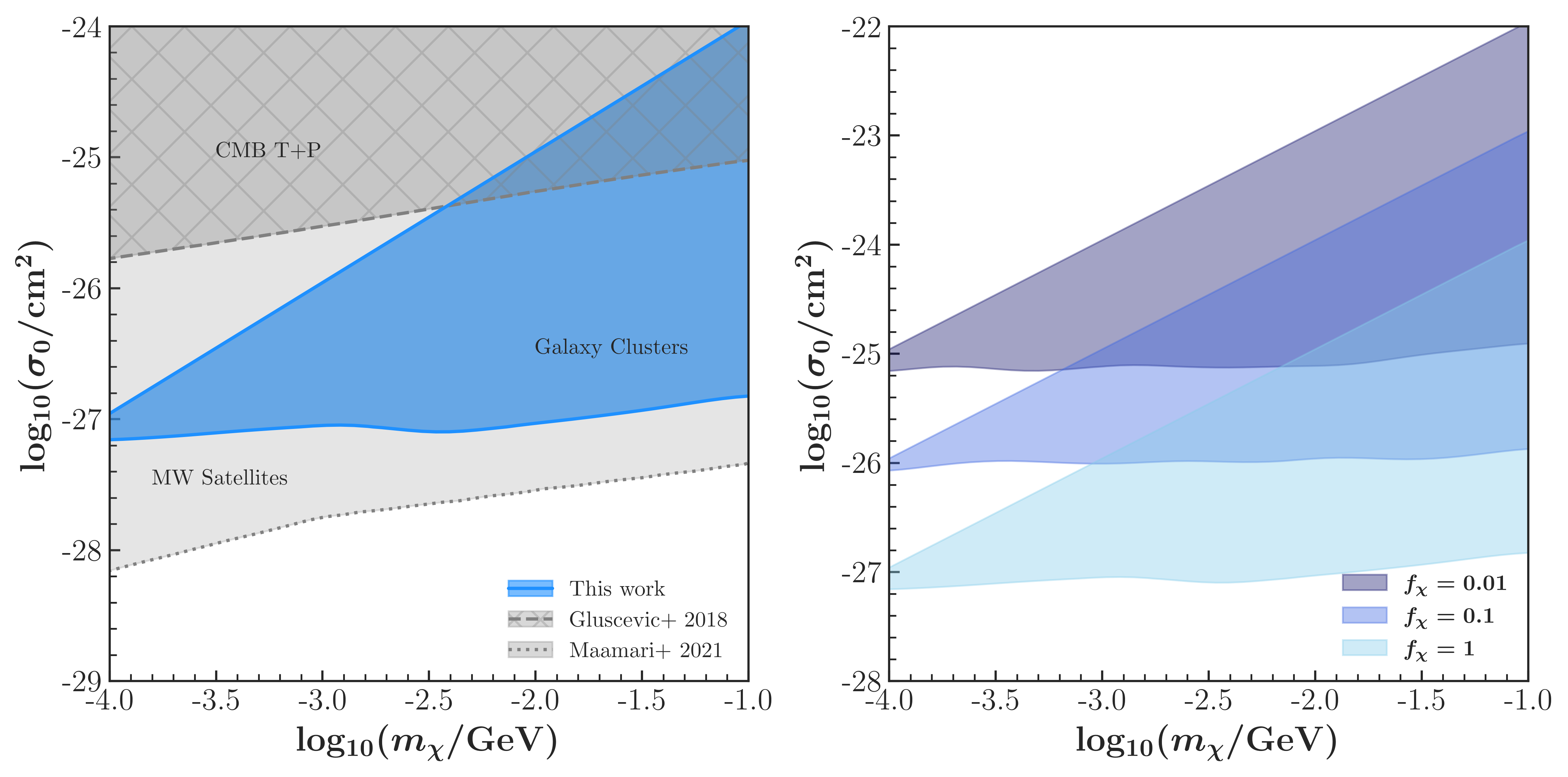}
    \caption{
\textbf{Left:} Upper bounds on the DM-proton velocity independent scattering cross section $\sigma_0$ as a function of $m_\chi$. The blue shaded region indicates the region excluded by this work, which is compared with constraints from MW satellites \citep{Maamari2021} (in gray with no hatching) and the CMB \citep{Gluscevic2018} (in gray with crossed diagonal lines hatching). The region above the blue shaded region represents the area excluded from the MCMC prior based on the interaction rate limit outlined in the end of section \ref{sec:modeling}. Within this region, the interaction rate is high enough that the DM and baryon temperatures can no longer be considered constant, and therefore this method cannot provide constraints within that region. We report orders of magnitude improvement over the ref.~\cite{Gluscevic2018} constraints, and comparable bounds to the ref.~\cite{Maamari2021} constraints within this mass range.
    \textbf{Right:} 95\% upper bounds on the DM-proton velocity independent scattering cross section for a 2-component scenario where part of the DM is non-interacting. $f_\chi$ represents the fraction of DM that has interactions with protons. The shaded regions represent the regions excluded by this work for different $f_\chi$. The strength of this constraint scales linearly with $f_\chi$.
    }
    \label{fig:constraints comparison}
\end{figure*}

\edit{We perform the MCMC analysis described in section \ref{sec:data} on the set of 49 clusters described above, and present details of those results. Running this analysis on the set of 111 virialized clusters, or the full set of 173 clusters with $z<0.2$ yields constraints within the same order of magnitude. With the parameters described above, the MCMC chains run on the 49 cluster set have a Gelman-Rubin statistic $\hat{R}=1.0012$ for $\log_{10}\sigma_0$ and $\log_{10}m_\chi$, indicating convergence. The same MCMC analysis done on the 111 and 173 cluster sets gives similar $\hat{R}$ values. }

This yields a 95\% upper bound of $\sigma_0<9.3\times10^{-28} \mathrm{~cm^2}$ for $m_\chi$ within the mass range $10^{-4} - 10^{-1}$ GeV. The resulting DM cooling rate is 1-4\% the magnitude of the radiative cooling rate for each cluster. \edit{The left panel of figure~\ref{fig:constraints comparison} shows the 95\% contours for $\sigma_0$ and $m_\chi$ obtained from MCMC in the 100\% interacting DM case for $n=0$ (velocity independent) interactions. The blue shaded region is ruled out by this model,} because within this area of parameter space the cooling rate due to DM-baryon interactions would be too high and the DM and baryons would be thermalized. The results are consistent with a model where there are no DM-baryon interactions. The constraints are approximately independent of mass within the mass range studied, because $m_\chi << m_p$ and $T_b>>T_\chi$ (see eq.~\ref{sigma_0 expression1}). These results are consistent within an order of magnitude for the sets of 111 and 173 clusters described in section \ref{sec:data}.

As an extreme and conservative case, we also consider a situation where the total energy in a cluster is transferred to DM through DM-baryon interactions over a cluster's lifetime. This is represented as
\begin{equation} \label{eqn total energy constraint}
    \frac{E_{\mathrm{tot}}}{t_{\mathrm{age}}} \geq \frac{dQ_{b\rightarrow\chi}}{dt}. 
\end{equation}
Here, $E_{tot}$ is the sum of the total potential energy of the cluster and the kinetic energy in baryons and DM. This expression is therefore independent of the modeling of other thermodynamic processes in clusters. From applying this expression to the set of 49 clusters, we obtain a 95\% upper bound of $\sigma_0=1.7\times10^{-24} \mathrm{~cm^2}$ for $m_\chi=10^{-4}$ GeV, and $\sigma_0=2.2\times10^{-24} \mathrm{~cm^2}$ for $m_\chi=10^{-1}$ GeV. These results represent that even in this extreme case where all energy in a cluster is transferred to DM through DM-baryon interactions, this method shows potential for constraining $\sigma_0$, regardless of how other heating and cooling mechanisms in the cluster are  modeled. For example, even if the heating rate from AGN feedback well exceeded the radiative cooling rate, this method would still place some constraints.

Incorporating the surface pressure term presented by ref.~\citep{Shaw2006} into the virialization step of the dataset selection changes the set of clusters used in our analysis. Defining how virialized a cluster is as $f_{vir} = \frac{2E_{th}-E_{sp}}{E_p}$ and keeping all clusters with $f_{vir}$ within 2 standard deviations of 0 leaves 55 clusters. If we perform the additional cut that requires the AGN heating rate for these clusters to be within 2 standard deviations of the radiative cooling rate, we get a final set of 44 clusters. Performing the MCMC analysis presented in this work on either of these sets of clusters yields constraints on $\sigma_0$ that are within an order of magnitude of the results presented in this paper.

The constraints in this work are also applicable to the fractional interacting DM (IDM) case. The scaling can be seen from eq.~\ref{sigma_0 expression1}, where $\rho_\chi$ is the density of the interacting component of DM. Through this relationship the upper bound of $\sigma_0$ should be inversely proportional to the fraction of DM that is interacting. As shown in figure~\ref{fig:constraints comparison}, the results for fractional IDM are approximately independent of mass, and we obtain an average upper bound at $f_\chi=0.1$ of $\sigma_0<1.1\times 10^{-26}\mathrm{~cm}^2$, and at $f_\chi = 0.01$ of $\sigma_0<8.2 \times 10^{-26}\mathrm{~cm}^2$. For comparison, the constraints on IDM from the MW satellite population \citep{Maamari2021} are expected to degrade rapidly for smaller fractions of IDM. This is because the MW satellite constraints depend on the matter power spectrum. The data is only sensitive to suppression of power below 25\% of the matter power spectrum predicted by CDM, which corresponds to an IDM fraction above 0.5 \citep{An2024}. Therefore, the MW satellite method cannot constrain IDM fractions below 0.5, making the constraints from galaxy clusters the best  existing constraints for IDM models where IDM makes up $<$ 50\% of the DM.

\section{Discussion \& Conclusion}
\label{sec: discussion}

In this work, we develop a model to constrain the DM-baryon interaction cross section by examining thermal equilibrium in galaxy clusters. We consider the equilibrium between heating from AGN feedback and cooling due to bremsstrahlung emission and DM-baryon interactions. Using a sample of 49 virialized clusters from the REFLEX catalog, with mass estimates from ACT, we derive upper limits on the DM-proton interaction cross section, focusing on velocity-independent interactions. Through an MCMC analysis, we constrain the cross section $\sigma_0$ for DM particle masses $m_\chi < 0.1 \mathrm{~GeV}$. 

Previous work has placed stringent constraints on DM velocity independent interaction cross sections. Planck 2015 CMB data limits $\sigma_0<3 \times 10^{-26} \mathrm{~cm}^{2}$ for 1 MeV with 95\% confidence \citep{Gluscevic2018}, and measurements of the MW satellite population from the Dark Energy Survey and Pan-STAARS1 \citep{Drlica-Wagner2020} provide stronger constraints, limiting $\sigma_0<2.8\times10^{-28}\mathrm{~cm}^{2}$ for a 10 MeV DM particle with $95\%$ confidence \citep{Maamari2021}. The limit obtained in this paper is within an order of magnitude of the ref.~\cite{Maamari2021} result in the mass range $m_\chi \sim 10^{-4} - 10^{-1}$ GeV, validating these results by providing a completely independent constraint. Additionally, this constraint is 1-2 orders of magnitude stronger than the CMB limit.

Ref.~\cite{Wadekar2021} examined the heating/cooling rate due to DM interacting with gas in the Leo T dwarf galaxy. Leo T is assumed to be stable over timescales associated with its astrophysical radiative cooling rate. For a system to be in a steady state, the magnitude of DM heating/cooling should be equal to the magnitude of the net heating/cooling due to other mechanisms in the galaxy. They consider the radiative cooling rate, stating that the DM heating/cooling should be lower in magnitude than the radiative cooling rate. By contrast, in this work we consider galaxy clusters that we expect to be in thermodynamic equilibrium due to their redshift and virialization, and develop a constraint on how strong DM cooling could be based on both AGN heating and bremsstrahlung, the major heating and cooling mechanisms of galaxy clusters. Ref.~\cite{Shoji2024} use a similar thermal equilibrium statement to ours to examine the impact DM-baryon interactions could have on the ICM. Their work focuses on the possibility of DM-baryon scattering to provide an additional heating source in cool core clusters. Our results are not in tension with theirs because we look at a parameter space where DM-baryon interactions result in cooling instead of heating.

AGN release huge amounts of energy into their surrounding environments, heating baryons through radiative processes, or mechanical processes such as jets and winds. Modeling this AGN feedback is an active area of research (the introduction of ref.~\citealp{Soker2022} gives a summary of many different AGN feedback mechanisms), with several models and simulations available to attempt to explain it. In this paper, we explore a mechanical effervescent heating model by ref.~\cite{Iqbal2023}, but different AGN heating models could predict different AGN heating rates, affecting the predicted $\sigma_0$ as in eq.~\ref{sigma_0 expression1}. The method outlined in this paper assumes that $T_b$ and $T_\chi$ remain at different temperatures over the course of a cluster’s lifetime. Therefore, the interaction rate limits the parameter space within which this model can be applied (see the end of section~\ref{sec:modeling}). AGN models that predict significantly more heating than the ref.~\cite{Iqbal2023} model would be outside of this allowed parameter space, as maintaining equilibrium in these clusters would require DM cooling to occur at a rate that would lead to thermalization of DM and baryons, after which point there would be no more DM cooling.

In addition to AGN feedback and radiative cooling mechanisms, there are other thermodynamic effects such as shock heating due to infalling material, and line emission which could be added to the radiative cooling rate due to bremsstrahlung in eq.~\ref{eqn cooling rate} \citep{Tozzi2001}. A detailed accounting of these processes is outside the scope of this work. It is worth noting that for clusters at low redshift and in equilibrium, we do not expect major mergers would cause significant shock heating and disturb the equilibrium \citep{Sereno2021}. Clusters at the redshifts we consider accrete mass more slowly \citep{Correa2015}, with major mergers, which would disturb equilibrium and cause significant heating, being rare. In addition, line emission is likely to be less efficient than bremsstrahlung. Changes to the heating and cooling models that lead to substantially increasing the net heating of the clusters before DM cooling is added will lead to predicted $\sigma_0$ that are outside of the allowed prior range due to interaction rates being too high, making the assumption that the DM and baryon temperatures are different and approximately constant invalid in those scenarios.

For some clusters, thermal and kinetic SZ data is available \citep{Amodeo2021}. These measurements provide more precise temperature and density profiles than we get from the hydrostatic equilibrium equation and temperature-luminosity scaling relations. SZ measurements also give information about the pressure profiles of clusters, which could constrain feedback models. Future work that combines SZ measurements of cluster profiles with more detailed heating and cooling models could lead to much stronger and robust constraints. With the upcoming Simons Observatory \citep{Ade2019}, this is likely to be possible in the near future.

This paper presents a novel method for deriving constraints on the DM-baryon interaction cross section and DM particle mass by considering the thermodynamics of galaxy clusters and the cooling rate implied by DM-baryon interactions.
This work examines DM particle masses $10^{-4}$ GeV $\leq m_\chi\leq$ $10^{-1}$ GeV, and reports a mass independent 95\% upper limit on the interaction cross section for velocity independent interactions of $\sigma_0<9.3\times10^{-28} \mathrm{~cm^2}$, as well as limits on $\sigma_0$ for models that are 1\% and 10\% IDM. By examining a simplified model of a galaxy cluster, we support existing bounds from MW satellites. Future work could involve implementing more thermodynamic processes, studying different AGN models, examining how the net heating rates of clusters change over their lifetimes, incorporating SZ temperature and density profiles, and using simulations for more accurate modeling of galaxy cluster thermodynamics.

The code used to generate the plots in this paper is available online at \url{https://github.com/eleanorstuart/thermo-idm}.

\acknowledgments
The authors wish to thank Vera Gluscevic, Yacine Ali-Ha\"imoud, John Wallin, and Elena Pierpaoli for comments on the manuscript and useful discussions. We also thank Srini Raghunathan, James Bullock, Manuel Buen Abad, Dimple Saarnaik, and Trey Driskell, and Wendy Crumrine for useful discussions about this work. 

\appendix

\section{AGN heating model} \label{app:agn model}
We use the effervescent heating model from ref.~\cite{Iqbal2023}, which is given in eq. \ref{eqn heating rate}
where

\begin{equation}
\label{h(r)}
    h(r)=\frac{L_{\mathrm{inj}}}{4\pi r^2}[1-\mathrm{exp}(-r/r_0)]\mathrm{exp}(-r/r_{\mathrm{cutoff}}) q^{-1}.
\end{equation}

In eq.~\ref{h(r)}, $L_{\mathrm{inj}}$ is the energy injection rate, $r_0 = 0.015R_{500}$, $r_{\mathrm{cutoff}}$ is the heating cutoff radius, and $q$ is a normalization factor related to the pressure profile (defined in \citealp{Iqbal2023}). The heating rate per volume $\epsilon_{\mathrm{heat}}$ is integrated over volume to obtain the total heating rate. 
$L_{\mathrm{inj}}$ is approximated using this relation derived in ref.~\cite{Iqbal2023}:
\begin{equation}
\label{eqn Linj-Mbh}
    L_{\mathrm{inj}} \approx \frac{10^{44}~\mathrm{erg/s}~M_{\mathrm{BH}} }{10^{9.5} M_\odot}. 
\end{equation}
We obtain $M_{\mathrm{BH}}$ using an $L_{500} -M_{\mathrm{BH}}$ relation \citep{Gaspari2019}: 
\begin{equation}\label{eqn-l500mbh}
\begin{split}
    \mathrm{log}(M_{\mathrm{BH}}/M_\odot) =  10 \pm 0.11 + (0.38 \pm 0.03) \mathrm{log}  (L_{500}/10^{44}\mathrm{~erg/s}).
\end{split}
\end{equation}
In this model, $L_{\mathrm{inj}}$ determines the overall magnitude of the heating rate. By deriving this value from $M_{\mathrm{BH}}$ we avoid using an AGN feedback model that is coupled to the radiative cooling rate.

\section{DM-baryon heat exchange} \label{app:dm-baryon}
The change in energy of a single baryon during a collision with a DM particle is $\Delta E_b = m_b \mathbf{v}_{cm} \cdot \Delta \mathbf{v}_b$, where $\mathbf{v}_{cm} = (m_b\mathbf{v}_b + m_\chi \mathbf{v}_\chi)/(m_b+m_\chi)$ is the velocity of the center of mass of the system, $\mathbf{v}_b$ is the velocity of the baryon and $\mathbf{v}_\chi$ is the velocity of the DM particle \citep{Munoz2015, Dvorkin2014, Boddy2018}. The change in velocity of the baryon is obtained as 
\begin{equation}
    \Delta \mathbf{v}_b = \frac{m_\chi}{m_b + m_\chi} |\mathbf{v}_b - \mathbf{v}_\chi|\left(\hat{n} - \frac{(\mathbf{v}_b - \mathbf{v}_\chi)}{|\mathbf{v}_b - \mathbf{v}_\chi|} \right), 
\end{equation}
where $\hat{n}$ is the final direction of the baryon after the collision. 
To obtain the total heat exchange rate for the entire cluster, this expression needs to be integrated over the velocity distributions of baryons and DM, over the interaction rate in all directions, and over the entire volume of the cluster. 
\begin{equation}
\begin{split}
     \frac{dQ_{b\rightarrow\chi}}{dt} = \int dV \rho_b \int d^3 v_\chi f_\chi \int d^3 v_b f_b   \\
    \int d\hat{n} \frac{d\sigma}{d\hat{n}} |\mathbf{v}_b - \mathbf{v}_\chi|n_\chi  \mathbf{v}_{cm} \cdot \Delta \mathbf{v}_b
\end{split}
\end{equation}

Plugging in $\Delta \mathbf{v}_b$ and $\mathbf{v}_{cm}$ and integrating over interaction rate and direction this becomes 
\begin{equation} \label{plug in vb}
\begin{split}
        \frac{dQ_{b\rightarrow\chi}}{dt} = \int dV \frac{\rho_b \rho_\chi \sigma_0}{(m_b + m_\chi)^2} \int d^3 v_\chi f_\chi 
        \\ \int d^3 v_b f_b |\mathbf{v}_b - \mathbf{v}_\chi| (m_b \mathbf{v}_b + m_\chi \mathbf{v}_\chi) \cdot (\mathbf{v}_b - \mathbf{v}_\chi)
\end{split}
\end{equation} 

 A substitution is made in order to solve the velocity distribution integrals \citep{Munoz2015}:
 \begin{equation}
     \mathbf{v}_m = \frac{\frac{m_\chi}{T_\chi}\mathbf{v}_\chi + \frac{m_b}{T_b}\mathbf{v}_b}{\frac{m_\chi}{T_\chi} + \frac{m_b}{T_b}}
 \end{equation}
 \begin{equation}
     \mathbf{v}_{th} = \mathbf{v}_\chi - \mathbf{v}_b
 \end{equation}. 

 Nothing depends on $\mathbf{v}_m$ so eq.~\ref{plug in vb} becomes
 \begin{equation} \label{substitution}
    \frac{dQ_{b\rightarrow\chi}}{dt} = \int dV \frac{\rho_b \rho_\chi \sigma_0(T_\chi - T_b)}{(m_b + m_\chi)^2u_{th}} \int d^3 v_{th} f_{th} (\mathbf{v}_{th}^3)
\end{equation} 

Integrating this expression for general $n$ is presented in ref.~\cite{Munoz2015}, and for the $n=0$ case we get
\begin{equation}
\label{final DM cooling}
        \frac{dQ_{b\rightarrow\chi}}{dt} = \int \frac{3(T_{\chi}-T_b)\rho_{\chi}\rho_b\sigma_{0}c_0 u_{th}}{(m_{\chi} + m_{b})^{2}} dV\; ,
\end{equation}
Where $c_0 = \frac{2^{5/2}\Gamma(3)}{3\sqrt{\pi}}$ and $u_{th}^2 = \frac{T_\chi}{m_\chi} + \frac{T_b}{m_b}$.









\end{document}